\documentclass{article}
\usepackage{spconf,amsmath,amssymb}
\usepackage{graphicx}
\usepackage{algorithm, algorithmic}
\usepackage{textcomp}
\usepackage{booktabs}
\usepackage{subcaption}
\usepackage{siunitx}
\usepackage{float}
\usepackage{hyperref}
\usepackage{xcolor}

\DeclareMathOperator*{\argmax}{arg\,max}

\makeatletter
\def\bstctlcite#1{\@bsphack\@for\@citeb:=#1\do{\edef\@citeb{\expandafter\@firstofone\@citeb}\if@filesw\immediate\write\@auxout{\string\citation{\@citeb}}\fi}\@esphack}
\makeatother

\title{Temporal Contrastive Alignment of EEG and Naturalistic Video Only Beats Linear Decoding On In-Sample Stimuli}
\name{Yuanyuan Yao$^1$, Simon Geirnaert$^{1,2}$, Alexander Bertrand$^{1}$\thanks{This research is funded by the Research Foundation - Flanders (FWO) project No G081722N, junior postdoctoral fellowship fundamental research of the FWO (for S. Geirnaert, No. 1242524N), Internal Funds KU Leuven (projects IDN/23/006, C14/25/108, and C3/25/107), and the Flemish Government (AI Research Program).\\All authors are also affiliated with Leuven.AI - KU Leuven institute for AI, Belgium.}}
\address{$^1$KU Leuven, Department of Electrical Engineering (ESAT), \\STADIUS Center for Dynamical Systems, Signal Processing and Data Analytics, Belgium\\
$^2$KU Leuven, Department of Neurosciences, Research Group ExpORL, Belgium\\
}

\begin{document}
\bstctlcite{BSTcontrol}
%
\maketitle
\begin{abstract}
\textit{Objective.} Neural tracking, the temporal alignment between (features of) a continuous stimulus and the elicited brain response, enables decoding attention from electroencephalography (EEG). Heavily studied for speech and music, it remains largely unexplored for naturalistic video, relying on linear canonical correlation analysis (CCA). We investigate whether a nonlinear temporal contrastive model can strengthen EEG-video alignment, and examine the physiological and practical basis of what it learns.
\textit{Approach.} We pair a lightweight EEG encoder with a projection of VideoMAE features, training both branches via a correlation-based InfoNCE objective. The model is benchmarked against CCA under three generalization regimes. To probe the basis and utility of the learned representations, we introduce a sham control with circularly shifted video features, a zero-shot transfer test to selective attention decoding on superimposed videos, and an evaluation against a group-averaged EEG reference.
\textit{Main results.} The model outperforms CCA on unseen subjects, matches CCA on held-out segments of seen videos, but falls behind on entirely novel videos. A sham control shows that this high accuracy on unseen subjects does not certify physically meaningful alignment: a model retrained on circularly shifted video features, with genuine temporal alignment destroyed, still reaches comparable accuracy on seen stimuli. Yet the coupling stays useful: on a zero-shot selective-attention decoding task with seen stimuli, the model still beats CCA, since decoding needs only a strong statistical coupling, regardless of whether it is physiologically meaningful. Finally, against a group-averaged EEG reference, genuine, generalizable coupling is nonetheless learned, only masked in single-subject EEG by measurement noise and subject-specific response differences.
\textit{Significance.} These results demonstrate that neural tracking models should be evaluated by holding out subjects and stimuli jointly. Progress toward open-set generalization hinges on data scale, stimulus diversity, and raising the single-trial signal-to-noise ratio.
\end{abstract}
%
\section{Introduction}
\label{sec:intro}

\par During the processing of continuous sensory stimuli such as natural speech or dynamic visual scenes, cortical neural activity synchronizes with the fluctuations of the input, creating a temporal alignment between the stimulus and the elicited electrophysiological response \cite{obleser2019neural, di2022neural}. This phenomenon, commonly known as neural tracking, is modulated by attention, where attention acts as an amplifier leading to stronger temporal coupling \cite{hillyard1998sensory, mesgarani2012selective, ki2016attention}. Because of this differential modulation, it is possible to decode the specific target of a person's attention and their overall level of attentional engagement directly from their brain activity. This activity is typically recorded using electroencephalography (EEG) due to its high temporal resolution and non-invasiveness. A key focus in recent years has been on translating proof-of-concept attention decoding into real-world applications, such as neuro-steered hearing aids and attention monitoring systems for clinical and educational settings \cite{di2022neural, geirnaert2021electroencephalography}.

\par However, such translation is non-trivial. Naturalistic stimuli, such as speech, music, or video, are rich and highly complex, posing significant challenges for finding relevant features that robustly correlate with the EEG response \cite{biesmans2016auditory, dmochowski2018extracting}. Moreover, to achieve true ecological validity, real-world applications require single-trial decoding. In these scenarios, the model must extract reliable information from inherently noisy EEG signals without the luxury of averaging across multiple trials \cite{geirnaert2021electroencephalography, o2015attentional}. Together, these constraints motivate investigating whether more expressive decoding models can form stronger temporal stimulus-EEG alignments and generalize well across diverse subjects and stimuli.

\par Extensive research in the speech domain has historically focused on linear methods, primarily linear stimulus reconstruction (LSR) and canonical correlation analysis (CCA). LSR learns a linear mapping from EEG to the stimulus \cite{biesmans2016auditory, o2015attentional, vanthornhout2018speech}, whereas CCA finds linear projections of both modalities that are maximally correlated \cite{de2018decoding, de2021auditory}. Although data-efficient and interpretable, these models have limited capacity to capture nonlinear relationships between EEG and complex naturalistic stimuli. An early step was taken by de Taillez et al.~\cite{de2020machine}, who extended the LSR framework with hidden layers, nonlinear activations, and dropout, representing a minimal, meaningful, but ultimately non-replicable departure from purely linear models \cite{geirnaert2021electroencephalography}. Ciccarelli et al.~\cite{ciccarelli2019comparison} benchmarked this nonlinear LSR variant alongside a convolutional neural network (CNN)-based model, showing that the CNN significantly outperforms both LSR and the nonlinear extension, a finding that was also non-replicable \cite{geirnaert2021electroencephalography}. Subsequent research explored a broader range of nonlinear architectures for EEG-stimulus alignment, including recurrent networks \cite{kuruvila2021extracting, lu2021auditory}, autoencoders \cite{chen2023auditory, bollens2022learning}, and transformers \cite{xu2022decoding, bollens2025contrastive}. In parallel, stimulus representations evolved from hand-crafted features (e.g., speech envelopes, Mel spectrograms) \cite{ciccarelli2019comparison, kuruvila2021extracting} to deep-learned semantic embeddings (e.g., Wav2Vec 2.0, HuBERT) \cite{thakkar2024investigating, hu2026effect}. Orthogonal to the choice of feature, the way the two modalities are related has also changed: instead of regressing one modality onto the other, two-branch models map EEG and stimulus into a shared embedding space with a trainable encoder on each path, jointly optimized by a match-mismatch \cite{monesi2020lstm, accou2021modeling} or, more recently, a contrastive \cite{chen2023auditory, bollens2025contrastive} objective, so that both representations are shaped by the alignment task itself. While these advances have improved decoding accuracy in certain conditions, their gains are fragile: higher-capacity models are more susceptible to exploiting dataset-specific shortcuts \cite{puffay2023relating, rotaru2024we, ivucic2024impact, heintz2026performance}, and performance often fails to generalize to new subjects, unseen stimuli, or different recording setups \cite{geirnaert2021electroencephalography}. Whether these improvements reflect genuine advances in stimulus-EEG alignment, or merely better in-distribution fitting, therefore remains an open question.

\par In the visual domain, the exploration of EEG alignment with naturalistic video stimuli remains limited. Early work by Dmochowski et al. \cite{dmochowski2018extracting} tested a variety of low-level video features and identified two (temporal contrast and optical flow) that could be correlated with EEG using CCA. However, Yao et al. \cite{yao2024identifying} later demonstrated that these significant correlations were primarily driven by shot cuts (sudden changes in scene or camera angle). These cuts introduce abrupt discontinuities that create artificial peaks in the extracted features while simultaneously eliciting strong visually evoked responses in the EEG. When tested on continuous, single-shot videos, these correlations became non-significant. To address this, the authors proposed object-based versions of the temporal contrast and optical flow features, successfully recovering significant correlations even in single-shot videos. To date, these remain the only features proven effective under such single-shot condition, and were subsequently utilized in a study applying CCA to decode attended videos in a superimposed-video paradigm \cite{yao2025eeg}. It is worth noting that a parallel line of recent research focuses on classifying or reconstructing visual stimuli from EEG using deep generative models \cite{du2023decoding, liu2024eeg2video, huang2026need}. However, these approaches typically operate on static images or short, discrete video clips. They rely heavily on decoding overarching semantic information to perform classification and reconstruction, which fundamentally differs from the continuous, stimulus-EEG temporal alignment problem focused on in this paper.

\par In this work, inspired by the advances in auditory EEG-stimulus alignment, we investigate for the first time whether introducing nonlinearity with neural networks and using more advanced deep video representations can improve temporal alignment between EEG and naturalistic video stimuli, compared to the CCA baseline. Aware of the pitfalls of data leakage and overfitting, the model is trained and evaluated under three generalization regimes: cross-subject (leave-one-subject-out), within-stimulus (leave-video-segment-out), and novel-stimulus (leave-one-video-out). For each regime, we dive into the underlying causes of success or failure, and identify the key constraints that limit performance. The paper is organized as follows. Section~\ref{sec:methods} presents the datasets, the pretrained video feature, the temporal contrastive model, and the evaluation protocol. Section~\ref{sec:results} benchmarks the model against CCA under the three generalization regimes and on a downstream attention-decoding task. Section~\ref{sec:discussion} examines what the model actually learns and what limits it, and Section~\ref{sec:conclusion} concludes.

\section{Methods}
\label{sec:methods}

\subsection{Data}

\par In this study, we worked with two publicly available continuous Video-EEG datasets collected at KU Leuven \cite{yao_2024_10512414, yao_2025_15665101}. Dataset I is our primary dataset for all model training and the three generalization regimes, while Dataset II provides a separate, held-out selective-attention test. These evaluation setups are detailed in Section~\ref{sec:evaluation} and Section~\ref{sec:attn_decoding}, respectively. In both datasets, participants watched naturalistic videos under free-viewing conditions, with no fixation constraint, to preserve ecological validity. The corresponding EEG data were acquired using a 64-channel BioSemi ActiveTwo system at a sampling rate of 2048 Hz. To record eye movements, four electrooculogram (EOG) electrodes were placed around the participants' eyes. The audio of all videos across both datasets was muted to prevent auditory-evoked neural responses from confounding the visual processing analysis.

\par In both datasets, the EEG was preprocessed following the same procedure as the studies that released them \cite{yao2024identifying,yao2025eeg}, comprising standard steps such as bad channel interpolation, rereferencing to the average of all channels, high-pass filtering with a cutoff frequency of 0.5 Hz, removing powerline noise at 50 Hz, and eye-artifact removal by regressing the EOG channels out of the EEG channels. For the present analysis, the preprocessed EEG was downsampled to 30~Hz for the CCA baseline, matching the video frame rate \cite{yao2024identifying,yao2025eeg}, and to 128~Hz for the contrastive model, which learns its own temporal filters and downsamples internally to the video-feature rate (Section~\ref{sec:eeg_branch}).

\subsubsection{Dataset~I: Video-EEG encoding-decoding dataset}
\par The first dataset \cite{yao_2024_10512414} was designed to investigate the temporal correlations between natural, single-shot videos and EEG signals during passive viewing. It includes two distinct subsets based on the visual complexity of the stimuli:
\begin{itemize}
  \item Single-Shot Subset: 19 participants passively viewed 13 continuous, single-shot videos (totaling approximately 63 minutes of data per participant). These videos depicted a single individual performing actions such as dancing, acrobatics, and mime.
  \item Complex Scene Subset: A subset of 9 participants viewed an additional 24-minute continuous clip from a television episode. This stimulus introduced more complex visual dynamics, including shot cuts, multiple moving objects, and complex interactions.
\end{itemize}
Since shot cuts introduce abrupt discontinuities that elicit strong event-related-potential-like responses in the EEG that may dominate the EEG-video coupling and overshadow the subtler neural tracking of the visual dynamics of interest \cite{yao2024identifying}, we focused our analysis on the single-shot subset (19 participants $\times$ 63 minutes of data).

\subsubsection{Dataset~II: Selective Visual Attention Decoding Dataset}
\par The second dataset~\cite{yao_2025_15665101} was designed to evaluate EEG-based decoding of selective visual attention when participants were presented with superimposed video streams. This dataset also includes data from 19 participants (a separate cohort from Dataset~I). The experimental paradigm used 14 single-shot videos (13 of which were inherited from Dataset~I), combining them into 7 superimposed video pairs. Each trial was structured to guide and test sustained selective attention:
\begin{itemize}
  \item Single-Object Phase (0-120s): Participants were presented with a single video stream and instructed to pay attention to the person in the video.
  \item Transition (120-122s): A two-second fade-in introduced a second, overlapping video stream.
  \item Superimposed-Object Phase (122s-End, 165s on average): Both videos were fully superimposed. Participants were instructed to maintain their visual attention solely on the original target from the first phase, actively ignoring the superimposed distractor. 
\end{itemize}
Each video pair was viewed twice with the target and distractor roles reversed, leading to a total of 14 trials per participant. The dataset comprises 19 participants $\times$ 28 minutes of single-object viewing and 19 participants $\times$ 38 minutes of superimposed viewing.

\subsection{Video Feature Extraction}

\par Representation learning for video is an extensively studied topic in computer vision, with a wide variety of architectures proposed in the literature. For neural tracking, the key requirement is a representation that captures temporal dynamics rather than one optimized primarily for semantic classification. Previous work established that motion-based features (object-based optical flow) are effective for EEG-video alignment \cite{yao2024identifying,yao2025eeg}, motivating a step toward richer spatiotemporal representations that go beyond a rough `all-pixel' average motion feature. To this end, we chose VideoMAE \cite{tong2022videomae}, a masked autoencoder architecture that learns representations by reconstructing largely masked video patches. The masking applies both spatially and temporally, forcing the model to learn spatiotemporal features that capture the underlying dynamics of the video, making it a natural fit for temporal alignment tasks. 

\par All videos were first standardized to a common format: resampled to 16 Hz and resized to $224 \times 224$ pixels. To preserve the original aspect ratio, frames were rescaled so that the longer side fit within 224 pixels and then symmetrically zero-padded to a square $224 \times 224$ canvas. These preprocessing choices match VideoMAE's original training configuration, which uses a 16-frame vision transformer base and an input resolution of $224 \times 224$. Each preprocessed video was partitioned into consecutive, non-overlapping 1-second windows of 16 frames. The model was loaded with a publicly released checkpoint on Hugging Face (MCG-NJU/videomae-base). Each input video segment yields 8 temporal token positions (16 frames / tubelet size 2) and $14 \times 14 = 196$ spatial patches per temporal position, for a total of 1568 tokens, each represented by a 768-dimensional embedding. The full hierarchy of representations of a 1-second video window includes the patch-embedding output together with the outputs of all 12 transformer blocks, giving 13 layers of hidden states: $(13, 8, 196, 768)$. The high dimensionality of this representation remains a challenge for downstream EEG alignment, so we applied a pooling step to reduce the dimensionality while preserving the temporal axis.

\subsubsection{Attention-weighted spatial pooling}
\label{sec:spatial_pooling}
\par
\begin{figure*}[t]
    \centering
    \includegraphics[width=\textwidth]{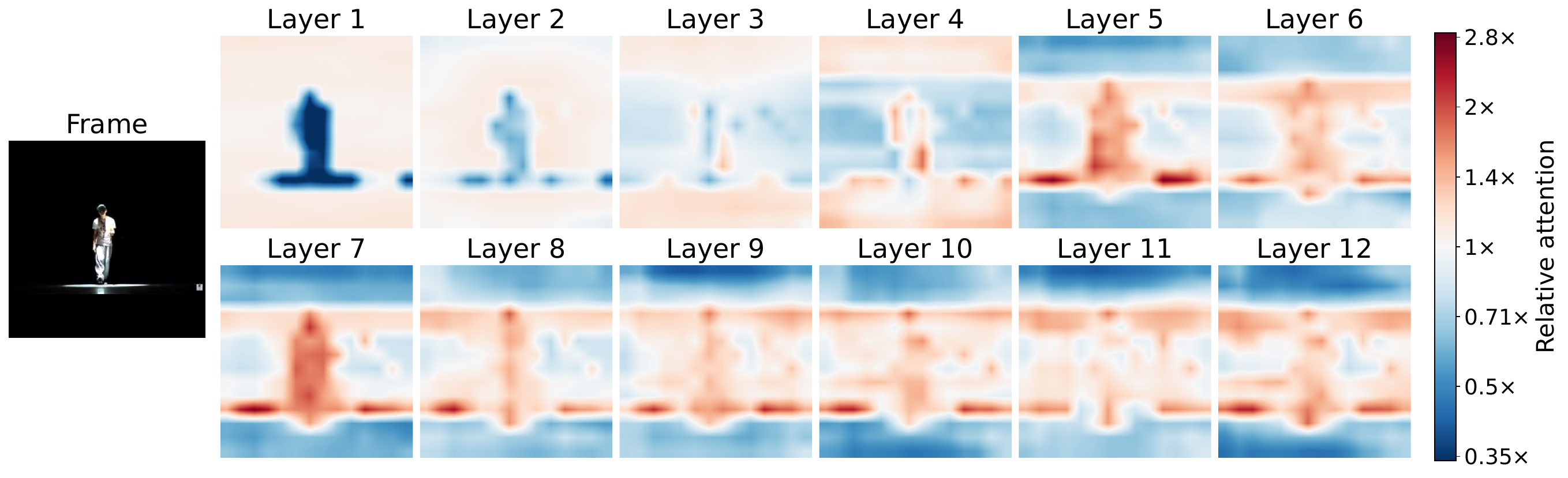}
    \caption{Layer-wise received attention across all 12 layers of VideoMAE for the first 1-second window of an example video. Left: the middle frame of the window, shown for reference. Right: received attention per spatial patch for each layer, averaged over heads, query tokens, and the 8 temporal positions, and upsampled from the $14 \times 14$ patch grid to the frame resolution. Colors show attention relative to a uniform baseline ($1\times$) on a shared $\log_2$ scale: blue represents below-uniform attention, and red represents above-uniform attention.}
    \label{fig:videomae_attention}
\end{figure*}

 We pooled spatially, layer by layer: for each of the 13 layers, we collapsed its 196 spatial patches into a single feature vector per temporal position, leaving the layer axis untouched. Rather than uniform spatial averaging, we used an attention-weighted pooling scheme driven by the model's own attention, so that spatially salient regions contribute more to the resulting representation. For a given layer, the multi-head attention tensor was averaged over heads, and for each token we computed the mean attention it receives across all query positions. These per-token saliencies (8 temporal $\times$ 196 spatial) were normalized with a softmax over the spatial dimension, which concentrates the pooling weights on the most attended regions. Figure~\ref{fig:videomae_attention} shows the attention maps of an example video for all 12 layers of the first window, next to the corresponding video frame, where the temporal axis is collapsed by averaging over the 8 positions (for display only). The attention maps reveal that early layers concentrate on the background and actively suppress the performer. This pattern inverts around layer 4, where the model begins to favor the performer and high-contrast structures, such as the light reflections on the floor. We observed that this progression is consistent across different time windows and distinct videos, though the onset of this inversion occasionally shifts one layer later.

\par Within each layer, the hidden states were then reduced to a weighted sum over the 196 spatial patches using these weights, yielding one 768-dimensional vector per temporal position and per layer. The patch-embedding layer, which has no associated attention map, was pooled using the attention weights of the first transformer block as the nearest available proxy. After pooling, each 1-second window was represented by a feature tensor of shape $(13, 8, 768)$, that is, layers $\times$ temporal positions $\times$ features. Concatenating the features across a $T$-second video, we obtained a feature tensor of shape $(T, 13, 8, 768)$.

\subsubsection{Temporal differencing}
\par
\begin{figure*}[t]
    \centering
    \begin{subfigure}{0.495\textwidth}
        \includegraphics[width=\textwidth]{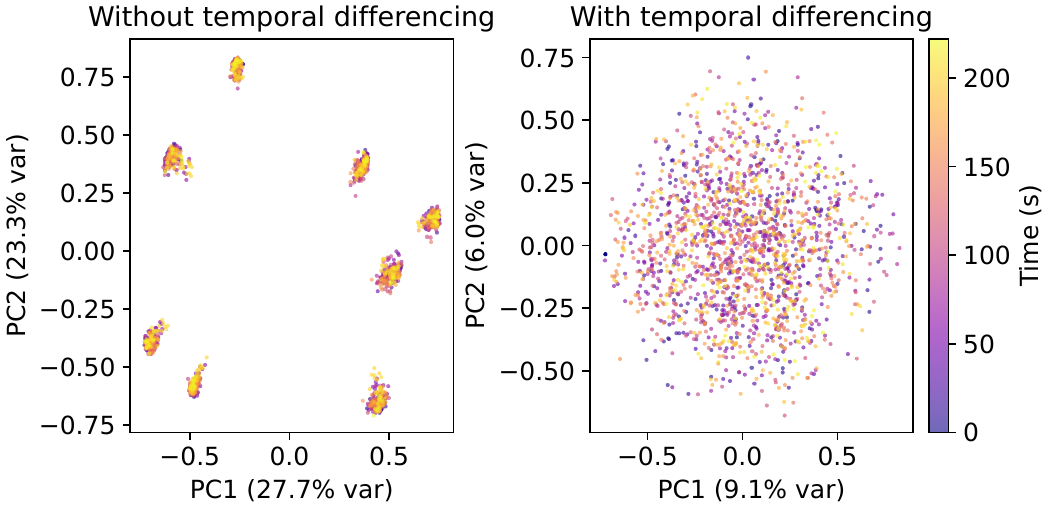}
        \caption{PCA scatter}
        \label{fig:raw_videomae_2D}
    \end{subfigure}
    \hfill
    \begin{subfigure}{0.495\textwidth}
        \includegraphics[width=\textwidth]{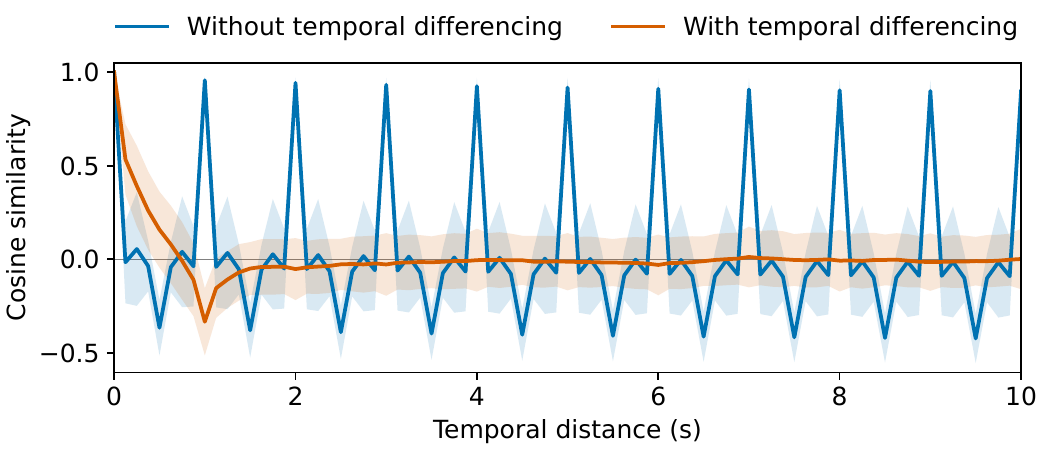}
        \caption{Normalized autocorrelation function}
        \label{fig:raw_videomae_decay}
    \end{subfigure}
    \caption{VideoMAE features of a representative video without and with temporal differencing. (a) PCA of the 8-Hz layer-12 features: each point is one time step projected onto the first two principal components (variance explained in the axis labels), colored by its time in the video. Without differencing (left), the features form eight clusters corresponding to the eight temporal positions within the 1-second window; with differencing (right), the clustering structure disappears. (b) Mean cosine similarity between feature vectors versus their temporal separation. Without differencing (blue), the similarity peaks at every integer-second lag; with differencing (orange), the periodic peaks disappear and the similarity decays to around zero.}
    \label{fig:raw_videomae}
\end{figure*}

 Before using these features for EEG alignment, we asked whether they primarily reflect the video's temporal dynamics or a static, position-dependent component. We inspected the pooled features of each video via principal component analysis (PCA): for each layer, we flattened the temporal axis to a single dimension of size $8T$ (for a single video) and projected the $768$-dimensional features onto their first two principal components. The resulting scatter plots revealed that the features of each video collapsed into eight tight clusters, consistent with the eight temporal positions within each 1-second window. Figure~\ref{fig:raw_videomae_2D} (left) shows one such case. To further characterize how feature similarity depends on temporal separation, we computed a normalized autocorrelation function from the $8T\times 8T$ pairwise cosine-similarity matrix comparing each token of a video with every other token. The normalized autocorrelation function is the mean similarity as a function of temporal lag $\Delta$, obtained by averaging the entries on the $\Delta$-th off-diagonal of the similarity matrix. The function (Figure~\ref{fig:raw_videomae_decay}, blue) shows a strong periodic pattern with period exactly one second: tokens at the same temporal position across different 1-second windows are highly similar. Moreover, this peak similarity shows only a marginal decrease as temporal distance increases. Both the clustering and the periodic autocorrelation were present in every transformer layer and in every video. This pattern is consistent with the influence of VideoMAE's positional embeddings: each window is encoded independently, but the token at a given temporal position carries the same fixed positional component in every window. This shared positional component, likely compounded by static background and semantic features, dominates the representation. Consequently, it drives a strong similarity between temporally aligned tokens, obscuring the genuine temporal variation we aim to study.

\par To remove this, we applied a first-order temporal difference across windows at each fixed temporal position, replacing each window's features by their difference from the preceding window:
\begin{equation}
\mathbf{f}'_{m} = \mathbf{f}_{m} - \mathbf{f}_{m-1} \quad \text{for } m = 2, \ldots, T,
\end{equation}
where $\mathbf{f}_{m}$ is the feature tensor of the $m$-th window with shape $(13, 8, 768)$, and $\mathbf{f}'_{m}$ is the differenced feature. Because the position-dependent offset is constant across windows for a given temporal slot, the subtraction cancels this constant component while preserving the dynamic component of the signal\footnote{Because VideoMAE uses joint space-time attention, token $i$ integrates information from the entire 1-second window, meaning the eight temporal slots do not strictly correspond to eight isolated 125~ms intervals. However, our differencing step does not assume this one-to-one correspondence. It solely requires the slot-specific nuisance component to be identical across windows, a condition guaranteed by the frozen encoder's repeated positional embeddings and empirically confirmed by Figure~\ref{fig:raw_videomae}.}. Figure~\ref{fig:raw_videomae} confirms that differencing removes the positional component. In the PCA (Figure~\ref{fig:raw_videomae_2D}, right), the eight clusters disappear, and the variance explained by the first two principal components drops from 51.0\% to 15.1\%. In the autocorrelation function (Figure~\ref{fig:raw_videomae_decay}, orange), the periodic peaks vanish, and the similarity instead decays to around zero. The resulting differenced features, flattened to an 8 Hz time series, have shape $(8T, 13, 768)$.

\subsection{Temporal Contrastive Model}

\par
\begin{figure*}[tp]
    \centering
    \includegraphics[width=1.9\columnwidth]{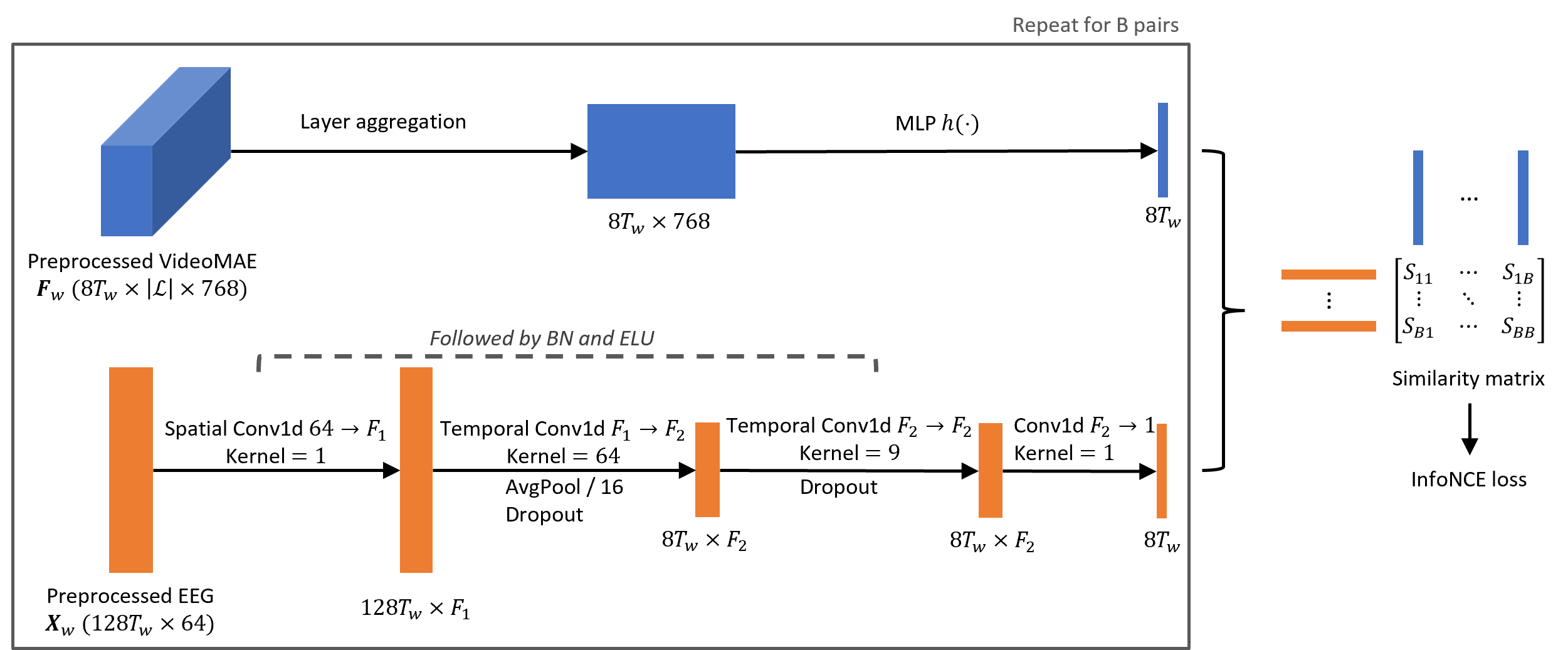}
    \caption{Architecture of the temporal contrastive model. The EEG branch (orange) applies spatial filtering followed by temporal convolutions to produce a scalar sequence at 8 Hz. The video branch (blue) aggregates pretrained VideoMAE features over the middle transformer layers and projects them to a scalar sequence via a lightweight MLP. Both outputs are zero-meaned and $L_2$-normalized such that their inner products become equivalent to a Pearson correlation, which populates the similarity matrix. The model is trained with a symmetric InfoNCE loss.}
    \label{fig:model_arch}
\end{figure*}

 To go beyond linear CCA while remaining tractable on a small dataset, we adopted a simple two-branch architecture trained with a contrastive objective. One branch processed $T_w$-second EEG windows; the other processed $T_w$-second video feature windows. Following CCA's use of temporal correlation as the similarity measure, both branches produced one scalar per temporal position, and similarity was computed as the Pearson correlation between the two output sequences. The overall architecture is illustrated in Figure~\ref{fig:model_arch}.

\subsubsection{Video branch}
\label{sec:video_branch}

\par Since spatiotemporal representation learning was delegated to the pretrained VideoMAE, the video branch consists of a lightweight projection head.

\par Denote the video features of a given window by $\mathbf F_{w} \in \mathbb{R}^{8T_w \times 13 \times 768}$. As observed in Figure~\ref{fig:videomae_attention}, the model transitions from background to foreground focus around layer 4. We specifically target these middle layers (transformer layers 4--7), as they capture rich, object-level information before the network's features become excessively abstract or specialized in the deepest layers. We therefore applied a learnable aggregation over $\mathcal{L}\in \{4,5,6,7\}$, the only step that mixes layers, to produce a single feature vector per temporal position:
\begin{equation}
\bar{\mathbf {F}}_{w}[t,:] = \sum_{\ell \in \mathcal{L}} a_\ell \mathbf F_{w}[t,\ell,:] \in \mathbb{R}^{d}, \quad t = 1,\dots,8T_w,
\end{equation}
where $a_\ell$ are softmax-normalized learnable weights over the layer dimension, and $d=768$ is the feature dimension. The aggregated features were then instance-normalized per window (each feature dimension zero-mean, unit-variance over the $8T_w$ timesteps) to account for scale variation in the pretrained features across videos, and passed through a multi-layer perceptron (MLP) $h$ applied at each timestep:
\begin{equation}
h(\cdot) = \mathrm{Linear}_{d \to d_h}\to\mathrm{GELU}\to \mathrm{IN} \to\mathrm{Linear}_{d_h \to 1},
\end{equation}
where $d_h$ is the projection width, $\mathrm{IN}$ denotes instance normalization over time, and dropout is applied before the final linear layer. The output of the video branch was a scalar sequence of length $8T_w$, denoted $\mathbf z^{\mathrm{vid}}_w$ with elements $z^{\mathrm{vid}}_t$ for $t=1,\dots,8T_w$:
\begin{equation}
\mathbf z^{\mathrm{vid}}_w = h\left(\mathrm{IN}(\bar{\mathbf{F}}_{w})\right).
\end{equation}
The only trainable parameters of the video branch were the layer aggregation weights $a_\ell$ and the MLP weights. The VideoMAE encoder was kept frozen throughout training. 

\subsubsection{EEG branch}
\label{sec:eeg_branch}

\par The EEG branch accepted 128 Hz EEG windows $\mathbf X_{w} \in \mathbb{R}^{128T_w \times C}$, where $C=64$ is the number of EEG channels. Inspired by EEGNet \cite{lawhern2018eegnet}, a spatial convolution first collapsed the channel dimension, mixing the $C$ electrodes at each time sample with $F_1$ learned spatial filters:
\begin{equation}
\underbrace{(128T_w,C)}_{\text{time}\times\text{channels}}
\xrightarrow{\text{spatial Conv1d }C\to F_1,\ k=1}\underbrace{(128T_w, F_1)}_{\text{time}\times\text{spatial filt.}}.
\end{equation}
With the channel axis collapsed, the remaining convolutions were temporal, operating along time. A 0.5-second kernel ($k = 64$) extracted temporal features, then average pooling with stride 16 downsampled the time axis from 128 Hz to 8 Hz, aligning EEG to the video time grid:
\begin{equation}
\underbrace{(128T_w, F_1)}_{\text{time}\times\text{spatial filt.}}\xrightarrow[\text{AvgPool }/16]{\text{temporal Conv1d }F_1\to F_2,\ k=64} \underbrace{(8T_w, F_2)}_{\text{time}\times\text{temporal filt.}}.
\end{equation}
A second filter ($k = 9$) then operated on this $8$ Hz grid to refine the temporal features, followed by a layer that mixed the $F_2$ filter responses into one scalar per timestep:
\begin{equation}
\underbrace{(8T_w, F_2)}_{\text{time}\times\text{temporal filt.}}\xrightarrow[\text{Conv1d }F_2\to 1,\ k=1]{\text{temporal Conv1d }F_2\to F_2,\ k=9}
\underbrace{(8T_w)}_{\text{scalar/step}}.
\end{equation}

\par All temporal convolutions used zero-padding of $\lfloor k/2 \rfloor$ to preserve the time axis, so the stride-16 average pooling mapped the 128 Hz sequence onto the same 8 Hz grid as the video features. Each convolution (except the final projection) was followed by batch normalization and ELU activation, with dropout after the pooling and after the second temporal filter. The EEG embedding output is denoted $\mathbf z^{\mathrm{eeg}}_w$.

\subsubsection{Contrastive loss}
\par Both branch outputs were zero-meaned and $L_2$-normalized over time, so their inner product is equivalent to the Pearson correlation. For a batch of $B$ matched pairs $\{(\mathbf z^{\mathrm{eeg}}_i, \mathbf z^{\mathrm{vid}}_i)\}_{i=1}^B$, define the similarity matrix
\begin{equation}
S_{ij} =\frac{\langle \mathbf z^{\mathrm{eeg}}_i, \mathbf z^{\mathrm{vid}}_j\rangle}{\tau} = \frac{\rho_{ij}}{\tau},
\label{eq:sim_mtx}
\end{equation}
with temperature $\tau$. The matched pairs lie on the diagonal, and we minimized the symmetric InfoNCE loss:
\begin{equation}
\text{Loss}
= -\frac{1}{2B}\sum_{i=1}^{B}
\left[
\log \frac{e^{S_{ii}}}{\sum_{j} e^{S_{ij}}}
+
\log \frac{e^{S_{ii}}}{\sum_{j} e^{S_{ji}}}
\right].
\end{equation}
This trains the matched EEG-video pair (on the diagonal) to score above all mismatched pairs in the batch, directly optimizing the match-mismatch discrimination used for evaluation (Section~\ref{sec:evaluation}).

\subsection{Training and Evaluation}

\subsubsection{Batch construction and window augmentation}
\label{sec:batch_and_aug}

\par We sampled $T_w$-second windows from the training set to form a batch of $B$ matched EEG-video pairs. We did not assign a fixed start position for each video, as this would create a discrete set of (video, window-index) identities. A model could then memorize a unique code for each such unit without learning any genuine temporal structure, a form of table lookup that would fail on unseen stimuli. Instead, the $B$ windows were drawn sequentially on-the-fly, with a running record of already-chosen start times per video. A candidate start $t_0$ was valid for a given video only if $|t_0 - t_0^*| \geq \Delta_{t}$ for every previously chosen start $t_0^*$ on that video, where $\Delta_t = \lceil 0.2T_w \rceil$. This capped the overlap between any two same-video windows in a batch, keeping all $B(B-1)$ off-diagonal similarities as unambiguous negatives. Start position was drawn uniformly within the available non-conflicting intervals of the chosen video. At each step the video was sampled proportional to the number of its remaining valid integer start positions, so videos with no room left dropped out automatically.

\par To encourage the model to learn representations that generalize across timescales, we applied a window augmentation strategy. For each batch, we randomly sampled a window length $T_w$ from a uniform distribution over $\{5, 10, 15, 20, 25, 30\}$ seconds. For the chosen (video, start, $T_w$), a subject was drawn uniformly from the training pool and the EEG window was sliced and z-scored per channel. The video features were sliced from the pre-extracted tensor at the same integer second boundaries. 

\par We will investigate three regimes, each with their own train, test, and validation split (see Section~\ref{sec:evaluation}). In validation, we used fixed $T_w = 30$-second windows extracted with a 5-second stride, to avoid the randomness of start positions affecting the evaluation. In test, we used the same 5-second stride but additionally evaluated the model on multiple window lengths to assess its robustness to timescale variation.

\subsubsection{Training}
\par Parameters were optimized using Adam, with learning rate set under a cosine-annealing schedule. The optimization hyperparameters were tuned once per evaluation regime and reserved validation set (Section~\ref{sec:evaluation}), not per test fold, using Optuna, choosing the best trial with the lowest validation loss out of 50 trials with epoch-level median pruning \cite{optuna_2019}. The search space is summarized in Table~\ref{tab:hparam_search}. Each trial was trained for up to 75 epochs of 60 batches each, with early stopping on the validation loss (patience of 10 epochs).

\begin{table*}[t]
  \centering
  \caption{Hyperparameter search space. Learning rate, weight decay, and
    temperature were sampled log-uniformly over the stated interval; the
    remaining parameters were sampled from the listed discrete sets. Tuning used
    Optuna (50 trials per study), minimizing the validation loss on a
    reserved held-out fold.}
  \label{tab:hparam_search}
  \begin{tabular}{@{}lll@{}}
    \toprule
    Hyperparameter & Role & Range / values \\
    \midrule
    Learning rate         & Adam step size                       & $[1\times10^{-4},\ 5\times10^{-3}]$, log-uniform sampling \\
    Weight decay          & Adam $L_2$ regularization            & $[1\times10^{-4},\ 1\times10^{-1}]$, log-uniform sampling \\
    Temperature $\tau$    & InfoNCE loss temperature             & $[0.05,\ 0.5]$, log-uniform sampling\\
    $F_1$                 & EEG encoder spatial filters              & $\{4, 8, 16\}$ \\
    $D$                   & EEG encoder depth multiplier, $F_2=F_1 \times D$              & $\{1, 2\}$ \\
    EEG dropout           & Dropout in the EEG encoder           & $\{0.3, 0.4, 0.5, 0.6, 0.7\}$ \\
    Projection hidden dim $d_h$ & Width of the video projection MLP    & $\{32, 64, 128\}$ \\
    Video dropout         & Dropout in the video projection head & $\{0.3, 0.4, 0.5, 0.6, 0.7\}$ \\
    Batch size            & Contrastive batch size               & $\{8, 16, 32, 64, 128\}$ \\
    \bottomrule
  \end{tabular}
\end{table*}

\subsubsection{Evaluation under three generalization regimes}
\label{sec:evaluation}

\par We separated three different generalization questions:
\begin{enumerate}
  \item Leave-one-subject-out (LOSO): Let $\mathcal{S}$ be the subject set ($|\mathcal{S}| = N_s$). Subject $s_0$ is reserved as the validation set. Each of the remaining $N_s - 1$ subjects is held out in turn as the test subject $s_i$, while the other $N_s - 2$ subjects are used for training. All videos appear in both splits. This scheme tests cross-subject generalization.
  \item Leave-video-segment-out (LVSO): Partition each video into $N_f = 8$ contiguous, non-overlapping segments. Fold $f_0$ of each video is reserved as the validation set, while each of the remaining $N_f - 1$ folds is held out in turn as the test fold $f_i$. All subjects and all videos appear in both splits. This regime tests within-manifold generalization, i.e., to unseen parts of seen videos.
  \item Leave-one-video-out (LOVO): Let $\mathcal{V}$ be the video set ($|\mathcal{V}| = N_v$). Video $v_0$ is reserved as the validation set. Each of the remaining $N_v - 1$ videos is held out in turn as the test video $v_i$, while the other $N_v - 2$ videos are used for training. All subjects appear in both splits. This regime tests off-manifold generalization, i.e., to unseen videos.
\end{enumerate}
The choice of the reserved subject $s_0$, fold $f_0$, and video $v_0$ affects the hyperparameter tuning and early stopping. Therefore, per regime we repeated the training and evaluation with three different reserved choices, which we will refer to as `hyperparameter configurations'. Results for each test subject, fold, or video were averaged over the three hyperparameter configurations in which it was tested. All three regimes operate exclusively on Dataset~I.

\par To evaluate the model, we used pairwise match-mismatch (MM) accuracy, a standard metric in neural tracking \cite{de2021auditory}. Given a video window and two EEG windows, one of which is the true match and the other an imposter, the model needs to identify which EEG window corresponds to the video. A dual formulation is to ask, given an EEG window and two video windows, which video is the true match. The model gives a similarity score for each pair, as defined in Equation~\ref{eq:sim_mtx}. The pairwise accuracy is the fraction of comparisons in which the matched similarity strictly exceeds the mismatched one. Formally:
\begin{equation}
\mathrm{acc}= \frac{\sum_{i}\sum_{j\neq i} \left[I(S_{ii} > S_{ij})+I(S_{ii} > S_{ji})\right]}{2B(B-1)},
\end{equation}
where $I(\cdot)$ is the indicator function. All windows in an evaluation batch, the true match and the imposters alike, were drawn from the held-out set, so no training data enters the comparison. The batch size was set to $B=8$ in validation and test, since LVSO's and LOVO's held-out sets are too short to accommodate larger batches. To give a robust estimate of the model's performance, we repeated the evaluation 5 times with different random seeds and reported the average accuracy.

\par The chance performance was estimated with a permutation test using the trained decoders. The EEG was circularly shifted relative to the stimulus features per video, so that the alignment between the two modalities was destroyed while preserving the temporal structure of each. The same evaluation procedure was applied to the permuted data. After 1000 permutations, the significance level is the 95th percentile of the null accuracies across folds and subjects, computed separately for each decoder and window length.

\subsubsection{Downstream selective attention decoding}
\label{sec:attn_decoding}

\par The three regimes above probe generalization within Dataset~I. To test whether the learned alignment also transfers to a qualitatively different task on a separate dataset, we additionally evaluated the model on the selective visual attention paradigm of Dataset~II, without updating any model parameters on that dataset. For this evaluation, a dedicated model was trained for match-mismatch discrimination on all 19 subjects of Dataset~I, with hyperparameters retuned for this setup, and then applied directly to attention decoding on Dataset~II. The single-object phase of Dataset~II served as the validation set for hyperparameter selection and early stopping. The superimposed-object phase was never seen during training or model selection and constitutes the true evaluation set.

\par During the superimposed-object phase, two video streams are presented simultaneously and fully overlaid. For a $T_w$-second EEG window $\mathbf{X}_w$, the model computes a similarity score against each competing stream:
\begin{equation}
s_k = \langle\mathbf{z}^{\mathrm{eeg}}_w,\, \mathbf{z}^{\mathrm{vid},(k)}_w\rangle, \quad k \in \{1, 2\},
\end{equation}
where $\mathbf{z}^{\mathrm{vid},(k)}_w$ is the normalized video branch output for stream $k$. The model predicts the attended stream as $\hat{k} = \argmax_k\, s_k$. Attention decoding accuracy is then the fraction of windows for which $\hat{k}$ matches the ground-truth attended stream.

\par Of the 14 videos in Dataset~II, 13 were inherited from Dataset~I. We excluded the one pair containing the single new video, leaving six pairs formed from twelve videos, all of which were present in the training set. Because each pair was viewed twice with the target and distractor roles reversed, every one of these videos appears equally often as the attended and as the unattended stream. The two competing streams in a trial are therefore equally represented in training, and the model holds templates of equal provenance for both. This is important for the interpretation in Section~\ref{sec:template_matching}: above-chance decoding cannot arise from the model scoring a better-learned video above a less familiar one, because neither stream is privileged. The evaluation is thus close to the LOSO regime, except that the task is now attention decoding, which tests whether the learned alignment is modulated by attention and decodable. The significance level was again estimated with a permutation test, this time by circularly permuting the EEG relative to both video streams.

\subsection{CCA baseline}

\par We compared the contrastive model against a CCA baseline from previous studies \cite{yao2024identifying, yao2025eeg}. In those studies, object-based optical-flow features were extracted from 30-Hz videos, the EEG was downsampled to 30 Hz, both modalities were hankelized to incorporate temporal information (15 lags spanning $-$467 to 0 ms for the video feature and 3 lags spanning $-$33 to 33 ms for each EEG channel), and were correlated using CCA. The sum of the first two canonical correlation coefficients served as the similarity score for the MM task. We followed the same procedure and parameter settings, with a single deviation: those studies fit a separate CCA per subject, whereas we fit one CCA on the EEG and video features of the subjects in the training split, concatenated along the time axis. To keep the results directly comparable, the CCA baseline was trained and tested under exactly the same three regimes as the contrastive model (Section~\ref{sec:evaluation}); the reserved validation set was excluded from fitting but served no further purpose, as CCA requires no hyperparameter tuning or early stopping. It was scored with the same pairwise accuracy metric, batch samplers, and seeds.

\section{Results}
\label{sec:results}

\subsection{Contrastive Model Outperforms CCA Under LOSO Without Exploiting Video Identity}
\label{sec:CL_wins_LOSO}

\par
\begin{figure}[ht]
    \centering
    \includegraphics[width=0.95\columnwidth]{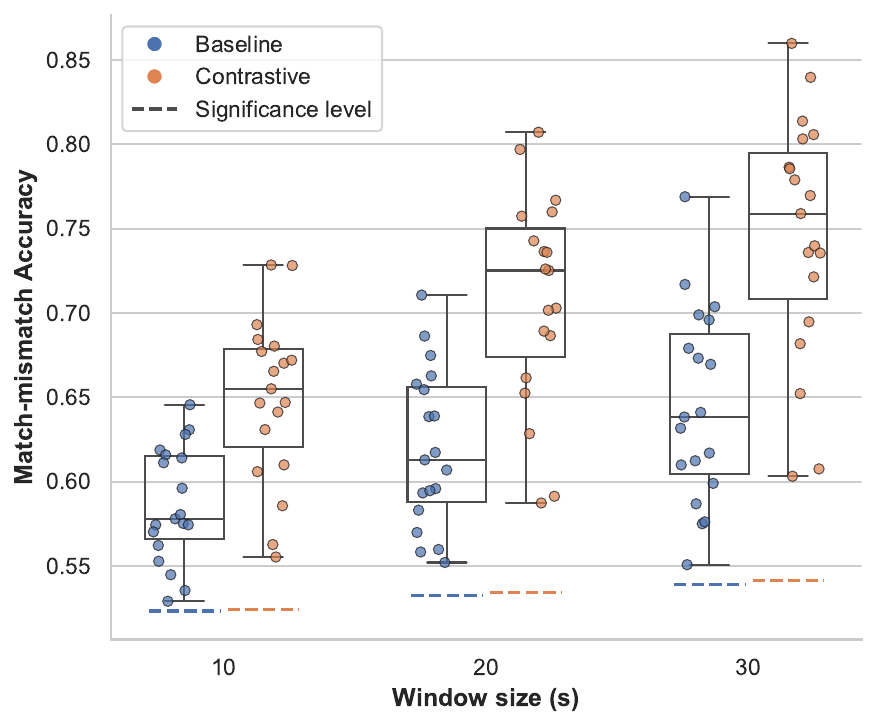}
    \caption{MM accuracy under LOSO for 10-, 20-, and 30-second windows: contrastive model (orange) vs.\ CCA baseline (blue). Each dot is one test subject averaged over hyperparameter configurations with different reserved validation subjects; boxes show the median and interquartile range (IQR), and whiskers extend to the most extreme data point within $1.5\times$IQR of the box edges. Dashed lines mark the significance level of each method.}
    \label{fig:loso_results}
\end{figure}

 Under LOSO, the contrastive model outperforms the CCA baseline across all tested window lengths. Figure~\ref{fig:loso_results} shows the MM accuracies of both methods, averaged over the three hyperparameter configurations with different reserved validation subjects. The mean accuracy of the contrastive model versus CCA is $65.0\%$ versus $58.6\%$, $70.8\%$ versus $61.9\%$, and $74.6\%$ versus $64.5\%$ for 10-, 20-, and 30-second windows respectively. Across the individual hyperparameter configurations, the mean accuracy gain of the contrastive model over CCA stays within $5.6\%\sim6.6\%$, $8.0\%\sim9.2\%$, and $9.1\%\sim10.5\%$ for 10-, 20-, and 30-second windows respectively. The narrowness of these ranges shows that the advantage is insensitive to which subject is held out as validation set for hyperparameter tuning. The improvements are statistically significant under a Wilcoxon signed-rank test across the 19 test subjects, with accuracies averaged over hyperparameter configurations ($p<0.001$ for all window lengths after Holm-Bonferroni correction).

\par
\begin{figure}[ht]
    \centering
    \includegraphics[width=0.95\columnwidth]{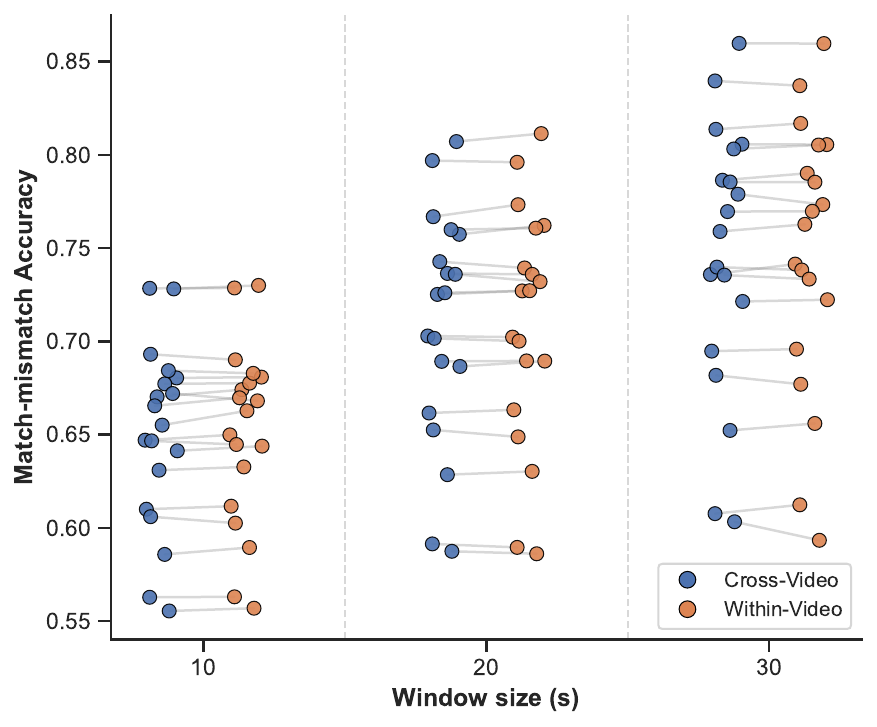}
    \caption{MM accuracy of the contrastive model under LOSO: standard cross-video evaluation (blue) vs.\ within-video evaluation (orange), where every imposter in a batch comes from the same video as the true match. Each connected pair of dots is one test subject, averaged over hyperparameter configurations with different reserved validation subjects.}
    \label{fig:loso_within_video}
\end{figure}

 Because all videos appear in both the training and test sets under LOSO, a natural concern is that the contrastive model achieves its advantage by exploiting video identity together with video-dependent EEG cues rather than genuine temporal alignment. In the standard evaluation, batches are formed without any constraint on which video each window comes from, so an imposter is usually (though not always) from a different video than the true match. A model could then separate matched from mismatched pairs using coarse video-level appearance cues in the EEG instead of moment-to-moment coupling. To rule this out, we performed a within-video evaluation in which all samples in a batch are drawn from the same video, so every imposter shares the same video as the true match and video identity carries no discriminative information. Figure~\ref{fig:loso_within_video} compares the within-video and standard evaluations, averaged over hyperparameter configurations. The accuracies barely change across subjects and window lengths, indicating that the model does not rely on shortcuts such as video identity, and this also holds within each hyperparameter configuration individually.

\par The within-video control rules out the coarsest shortcut, video identity, but not every non-temporal route to high LOSO accuracy. In Section~\ref{sec:sham_experiment} we probe a subtler, synthesized correspondence that can inflate LOSO performance without physically meaningful EEG-video coupling.

\subsection{The Advantage Disappears Under LVSO}
\label{sec:tie_LVSO}

\par
\begin{figure}[tp]
    \centering
    \includegraphics[width=0.95\columnwidth]{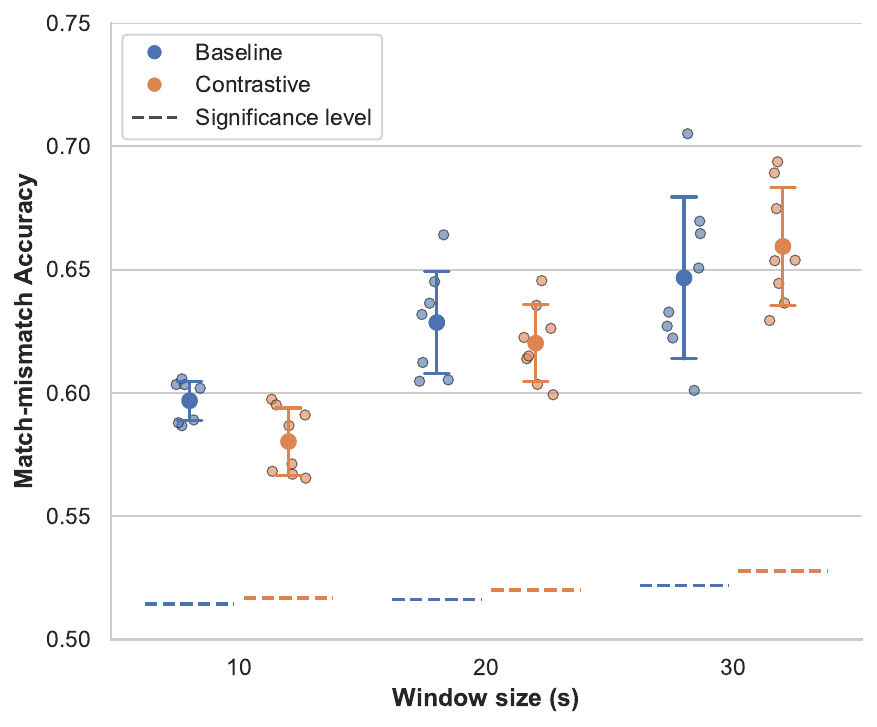}
    \caption{MM accuracy under LVSO for 10-, 20-, and 30-second windows: contrastive model (orange) vs.\ CCA baseline (blue). Each dot is one test fold averaged over hyperparameter configurations with different reserved validation folds; the large dot is the mean over folds and the error bars show $\pm1$ SD. Dashed lines mark the significance level of each method.}
    \label{fig:lvso_results}
\end{figure}

 Under LVSO, the contrastive model's advantage over CCA disappears. This regime holds out unseen temporal segments of videos that are otherwise present in training, so it probes on-manifold (interpolation) generalization, in contrast to the novel-video extrapolation tested under LOVO (Section~\ref{sec:CL_loses_LOVO}). Figure~\ref{fig:lvso_results} shows the MM accuracies of both methods, averaged over hyperparameter configurations with different reserved validation folds. The mean accuracy of the contrastive model versus CCA is $58.0\%$ versus $59.7\%$, $62.0\%$ versus $62.9\%$, and $65.9\%$ versus $64.7\%$ for 10-, 20-, and 30-second windows respectively. None of these differences is statistically significant under a Wilcoxon signed-rank test across the 8 test folds after Holm-Bonferroni correction ($p_{\mathrm{adj}} = 0.052, r = -0.94$ for 10~s; $p_{\mathrm{adj}} = 0.53, r = -0.39$ for 20~s; $p_{\mathrm{adj}} = 0.53, r = 0.44$ for 30~s, where $p_{\mathrm{adj}}$ is the Holm-Bonferroni-corrected p-value and $r$ is the rank-biserial correlation effect size), although at 10~s there is a trend in favor of CCA.

\par The differences are also small in magnitude. The contrastive model is more sensitive to the choice of validation fold: averaged over folds, the standard deviation (SD) of its accuracy across hyperparameter configurations is $1.2\%$, $2.1\%$, and $2.7\%$ for 10-, 20-, and 30-second windows, compared with at most $0.4\%$ for CCA, which involves no hyperparameter tuning or early stopping and varies only because the reserved fold is excluded from its training data. At 20 and 30~s, the difference between the two methods is smaller than this validation-induced variability, and at 10~s it is comparable. We therefore interpret the two methods as performing on par on unseen video segments, with no systematic advantage for either method.

\subsection{The Advantage Reverses Under LOVO}
\label{sec:CL_loses_LOVO}

\par
\begin{figure}[ht]
    \centering
    \includegraphics[width=0.95\columnwidth]{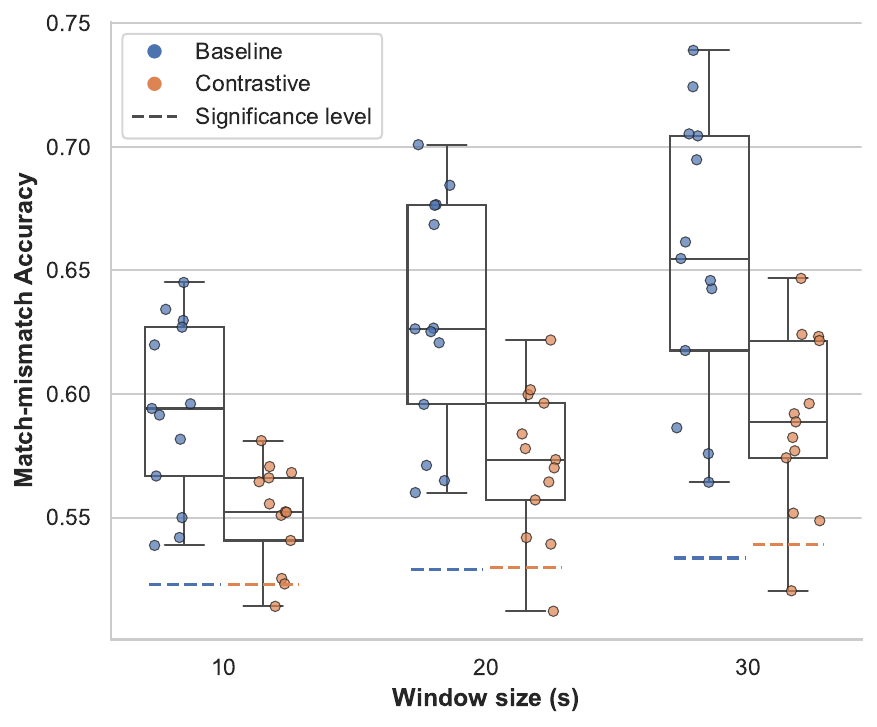}
    \caption{MM accuracy under LOVO for 10-, 20-, and 30-second windows: contrastive model (orange) vs.\ CCA baseline (blue). Each dot is one test video averaged over hyperparameter configurations with different reserved validation videos; boxes show the median and IQR, and whiskers extend to the most extreme data point within $1.5\times$IQR of the box edges. Dashed lines mark the significance level of each method.}
    \label{fig:lovo_results}
\end{figure}

 Under LOVO, which requires the model to generalize to entirely unseen videos (an off-manifold, extrapolation setting), the contrastive model's advantage reverses: it consistently underperforms the CCA baseline. Figure~\ref{fig:lovo_results} shows the MM accuracies of both methods, averaged over hyperparameter configurations with different reserved validation videos.
The mean accuracy of the contrastive model versus CCA is $55.1\%$ versus $59.4\%$, $57.2\%$ versus $63.1\%$, and $58.8\%$ versus $65.5\%$ for 10-, 20-, and 30-second windows respectively. The gap in favor of CCA is statistically significant under a one-sided Wilcoxon signed-rank test across the 13 test videos ($p_{\mathrm{adj}} = 0.002$ for all window lengths after Holm-Bonferroni correction). The reversal is consistent, but it does not imply that the model has learned no genuine EEG-video coupling. As we show in Section~\ref{sec:bottleneck}, the contrastive model still learns coupling that generalizes to unseen videos; what limits its single-trial accuracy on novel stimuli is individual noise\footnote{Throughout, \emph{individual noise} denotes everything in a subject's EEG that is not shared across subjects: both measurement noise and individual differences in how subjects respond to the video.}, not an absence of learned alignment.

\subsection{The Advantage Transfers to Downstream Attention Decoding on Seen Videos}

\par
\begin{figure}[ht]
    \centering
    \includegraphics[width=0.95\columnwidth]{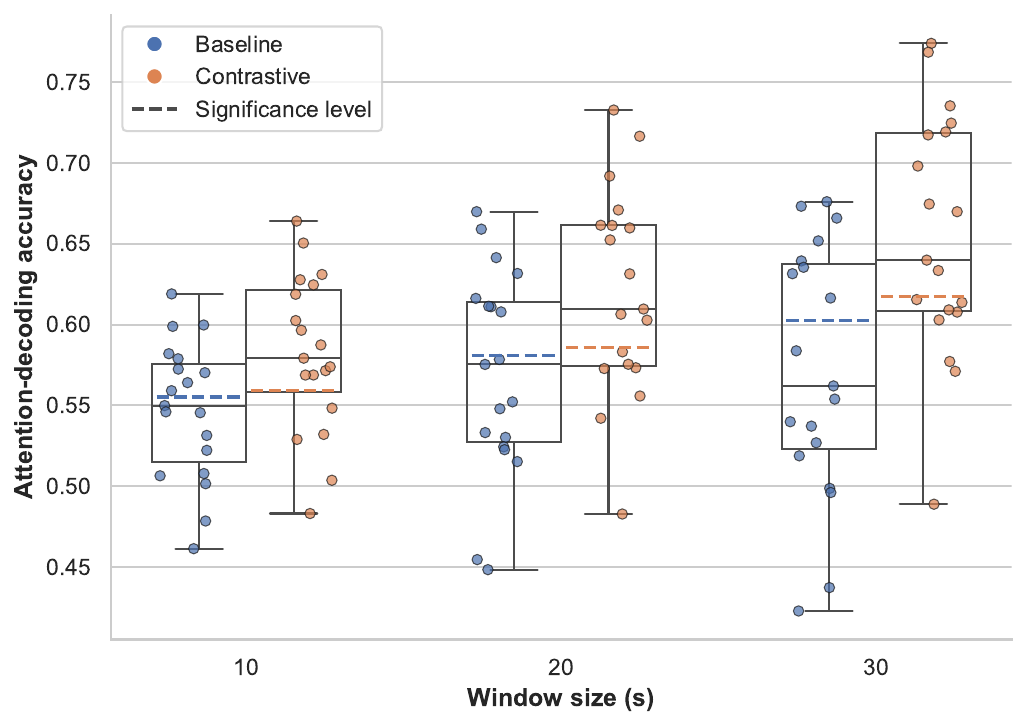}
    \caption{Attention-decoding accuracy on the superimposed-object phase of Dataset~II for 10-, 20-, and 30-second windows: contrastive model (orange) vs.\ CCA baseline (blue). Models are trained on Dataset~I and validated on the single-object phase of Dataset~II, without retraining. The one pair containing the video absent from Dataset~I is excluded, leaving six pairs formed from twelve videos, all seen during training and each serving equally often as the attended and the unattended stream. Each dot is one subject ($n=19$); boxes show the median and IQR, and whiskers extend to the most extreme data point within $1.5\times$IQR of the box edges. Dashed lines mark the significance level of each method.}
    \label{fig:attention_decoding}
\end{figure}

 Finally, we asked whether the learned alignment transfers to a qualitatively different, application-relevant task: decoding which of two simultaneously presented videos a subject attends to. As described in Section~\ref{sec:attn_decoding}, the model was trained on Dataset~I, validated on the single-object phase of Dataset~II, and tested on the superimposed-object phase of Dataset~II, without any retraining. Dataset~II adds one video to those in Dataset~I, and we excluded the video pair containing that new video from the evaluation, so all test videos were seen during training and this evaluation is close to the LOSO regime. It is nonetheless harder than MM: both video streams are simultaneously encoded in the EEG, and the model must identify the attended one, which is only possible if the learned EEG-video alignment is modulated by attention. The contrastive model outperforms CCA at all window lengths (Figure~\ref{fig:attention_decoding}): $p_{\mathrm{adj}} = 0.014, r = 0.64$ for 10~s, $p_{\mathrm{adj}} = 0.014, r = 0.62$ for 20~s, and $p_{\mathrm{adj}} = 0.002, r = 0.85$ for 30~s.\footnote{The CCA results are not directly comparable to the attention decoding results reported in \cite{yao2025eeg} on the same superimposed-video paradigm, where the decoders were subject-specific (a separate decoder fit per subject). Also, the significance testing applied here is more rigorous than in \cite{yao2025eeg}, as their approach artificially introduced independent segments when estimating the null distribution, resulting in an anti-conservative estimate.} The significance level is higher than in the MM task, because the attended versus unattended comparison is one-to-one, whereas in MM each matched pair is compared against multiple imposters. The larger effective sample size in MM concentrates the null accuracies near chance, lowering the significance threshold.

\section{Discussion}
\label{sec:discussion}

\subsection{High Accuracy Under LOSO Does Not Guarantee Physically Meaningful EEG-Video Alignment}
\label{sec:sham_experiment}

\par
\begin{figure}[ht]
    \centering
    \begin{subfigure}{0.45\textwidth}
        \includegraphics[width=\textwidth]{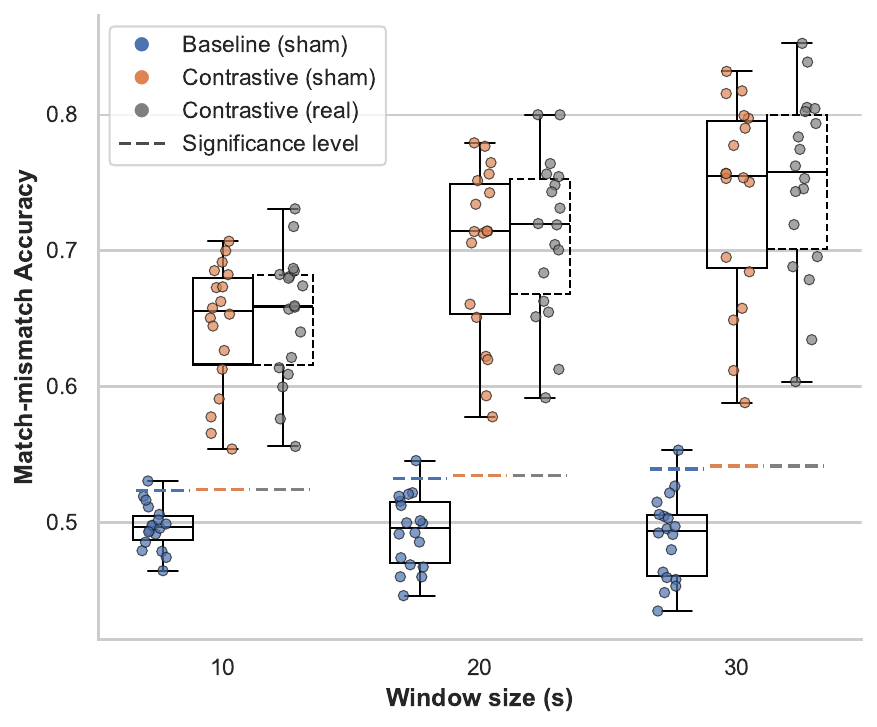}
        \caption{Sham vs. real accuracy}
        \label{fig:sham_vs_real}
    \end{subfigure}
    \hfill
    \begin{subfigure}{0.45\textwidth}
        \includegraphics[width=\textwidth]{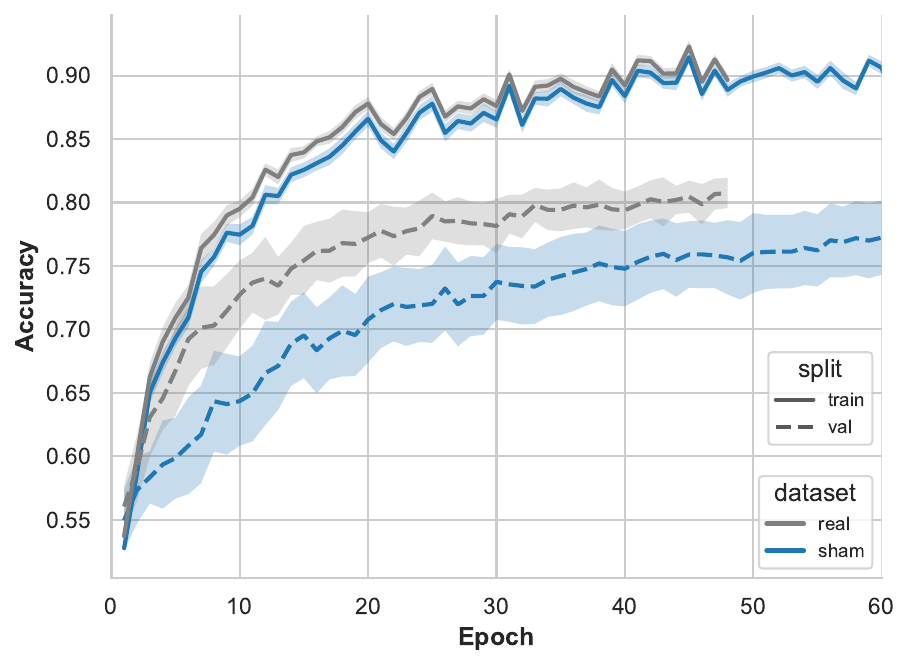}
        \caption{Training and validation curves}
        \label{fig:sham_training_curves}
    \end{subfigure}
    \caption{(a) MM accuracy under LOSO for 10-, 20-, and 30-second windows: contrastive model trained on real data (grey, dashed box) versus contrastive model (orange) and CCA baseline (blue) trained on sham data (per-video circularly-shifted video features). Each dot is one test subject; boxes show the median and IQR, whiskers extend to the most extreme data point within $1.5\times$IQR of the box edges, and colored dashed lines mark the significance level of each method. (b) Training (solid) and validation (dashed) accuracy versus epoch for the contrastive model trained on real (grey) and sham (blue) data, averaged over leave-one-subject-out folds; shading is $\pm1$ SD across folds.}
    \label{fig:sham_results}
\end{figure}

 In Section~\ref{sec:CL_wins_LOSO} we showed that the contrastive model outperforms CCA under LOSO without using video identity as a shortcut, and the window augmentation of Section~\ref{sec:batch_and_aug} further discourages the model from memorizing a discrete code for each (video, window-index) pair. To probe whether a more subtle form of shortcut exists, we constructed a sham dataset in which the genuine temporal alignment between EEG and video is destroyed while all other structure is preserved: we applied a per-video deterministic circular shift ($\geq 60$~s), the same one for all subjects, to the video features and left the EEG unchanged. The model was retrained on this sham dataset and evaluated under LOSO, reusing the hyperparameter configuration selected on real data with Subject~1 as the reserved validation subject. Architecture, learning rate, batch size, and schedule were therefore identical across the two conditions, which differ only in whether genuine EEG-video alignment is present in the training data. Surprisingly, the sham model reaches accuracy close to that of the corresponding real-data Subject~1 run (Figure~\ref{fig:sham_vs_real}), whereas the CCA baseline, given the same sham data, collapses to chance. The contrastive model can therefore reach high LOSO accuracy without genuine temporal alignment in the training pairs, through a route that is unavailable to CCA.

\par To see how this is possible, it helps to view the training set as a whole rather than as a collection of windows. Even though circular shift breaks the temporal alignment between EEG and video, the model may eventually build a complex, physically arbitrary mapping from the video-feature sequences to the corresponding EEG traces over the entire training set (i.e., across all subjects), which is then applied to each window. This is what window augmentation cannot resolve, because it only prevents the model from memorizing a fixed set of window slots, not from overfitting or memorizing a function that maps video content to EEG content. This function is unlikely to extrapolate to unseen videos, but it can interpolate to windows within the training videos, and, through the subject-general EEG response component, to EEGs of unseen subjects viewing the same videos. 
This mirrors the pattern across the three regimes: any stimulus-specific correspondence, whether genuine but confined to the seen video content or synthesized through a physically arbitrary mapping, is available wherever the model only needs to interpolate between different window sizes and window start-times within seen videos, inflating LOSO (Section~\ref{sec:CL_wins_LOSO}) and matching CCA under LVSO (Section~\ref{sec:tie_LVSO}), and it vanishes precisely where the model must extrapolate to novel videos (LOVO, Section~\ref{sec:CL_loses_LOVO}).

\par The sham experiment exemplifies a synthesized correspondence that can inflate LOSO accuracy without physically meaningful EEG-video coupling, but it does not necessarily imply that the model trained on real data also relies on such a nonsensical correspondence. Deep networks have the capacity to fit pure noise \cite{zhang2016understanding}; yet when genuine structure is present, they prioritize learning simple, generalizable patterns before resorting to memorization \cite{arpit2017closer, kalimeris2019sgd}. Here the physically meaningful coupling is plausibly simpler than the synthesized one. We therefore expect the model trained on real data to exploit the physically meaningful coupling whenever it is learnable. 

\par The learning dynamics are consistent with this hypothesis. Figure~\ref{fig:sham_training_curves} plots mean training and validation accuracy across LOSO folds against epoch. The training curves are nearly superimposed: both conditions plateau around $90\%$ (reaching $85\%$ at 17 versus 19 epochs)\footnote{The epoch-to-epoch fluctuations of the two training curves are highly correlated because the same random seed, shared across conditions and folds, determines the window length sampled at the start of each epoch, and window length dominates per-epoch training accuracy.}. The validation curves diverge: the real model reaches $70\%$ in 8 versus 19 epochs for the sham model, and $75\%$ in 15 versus 40 epochs, and its checkpoint is selected 10 epochs earlier on average. 

\par During training, the goal is to reduce the loss as much as possible, and the model will exploit whatever structure is available. Since it does not necessarily require genuine video-to-EEG coupling, the training curves are remarkably similar when training on real versus sham data. When transferring to an unseen subject, only the found patterns of the subject-independent, stimulus-locked EEG component carries over, which is what the validation curves track. This divergence suggests that the real model may exploit the simpler, physically meaningful coupling early in training, whereas the sham model must rely on a more complex, artificial transformation of the sham videos to these general EEG patterns, which takes longer to develop \cite{arpit2017closer, kalimeris2019sgd}.

\par In summary, the sham experiment shows that high accuracy under LOSO does not by itself certify that the model has learned physically meaningful EEG-video alignment: a reproducible but synthesized correspondence fit to the seen stimuli is sufficient to explain it. At the same time, the faster emergence of cross-subject transfer in the real model suggests that genuine coupling is present and is exploited when available, which will be further demonstrated in Section \ref{sec:bottleneck} with actual held-out video data. Part of what is learned may nonetheless be stimulus-specific, either genuine but confined to particular video content, or synthesized and built up in the later stages of training. 

\subsection{Physically Meaningful Coupling Is Not Always Required: Attention Decoding as Template Matching}
\label{sec:template_matching}

\par High LOSO accuracy may (at least partially) rest on a synthesized correspondence rather than (or in addition to) physically meaningful video-to-EEG coupling. What does this imply for the practical utility of the model? For applications that require generalization to unseen stimuli, the synthesized correspondence is insufficient, as shown in the LOVO setting. But many applications do not require generalization to unseen stimuli, only to unseen subjects. For example, a lecture video is fixed and known at training time, and the model is used to monitor attention of new students watching the same video. In this case, the model can operate as a template matcher: it learns a reproducible template for each video and matches incoming EEG against these templates. Whether the templates encode physically meaningful coupling and/or a synthesized correspondence to the EEG responses is irrelevant, as long as they are reproducible across subjects.

\par In the downstream attention decoding task of Section~\ref{sec:attn_decoding}, every evaluated video was seen during training, so the task is close to the LOSO regime, and each video serves equally often as the attended and as the unattended stream, so neither competing template is better learned than the other. Attention amplifies EEG responses to the attended stream, and therefore leads to a stronger match with the corresponding video template, regardless of whether the template reflects physically meaningful coupling or a synthesized correspondence. In other words, this is possible because the discriminating information lies on the EEG side rather than in the fidelity of the video mapping. Therefore, the advantage of the contrastive model over CCA on this task demonstrates practical utility and attention modulation, but, consistent with Section~\ref{sec:sham_experiment}, it does not by itself certify that the model has learned how to extract features from any video that are predictive for the EEG response.

\subsection{Generalizable Video-EEG Coupling Is Learned but Masked by Individual Noise}
\label{sec:bottleneck}

\par The LVSO tie and the LOVO reversal between CCA and the contrastive learning model raise a question: does the contrastive model fail to learn generalizable EEG-video coupling, or does it learn a generalizable coupling that is simply masked by individual noise in single-subject EEG? To separate these, we constructed a cleaner, group-averaged EEG reference, which suppresses individual noise because averaging cancels what is not shared across subjects. The regimes themselves are untouched: the model is fitted exactly as under LVSO or LOVO, so the held-out video segment or video is unseen during training for every subject, and the group-averaged reference enters only at evaluation time. When evaluating subject $n$ on a held-out segment or video, the group-averaged reference is the average of the EEG embeddings of the other $N_s-1$ subjects computed on that same held-out segment or video, time-aligned with the trial being scored.

\par With this reference, we defined two tasks analogous to the standard video-EEG MM task: the video-avgEEG task matches the video embedding against the group-averaged reference EEG embedding, and the EEG-avgEEG task matches subject $n$'s individual EEG embedding against the group-averaged reference (with subject $n$ removed). Because individual noise is largely averaged out of the reference, video-avgEEG estimates how well the learned alignment generalizes once individual noise is removed, and comparing it against EEG-avgEEG isolates whether the limitation lies on the video side or the individual-EEG side.

\par
\begin{figure}[ht]
    \centering
    \begin{subfigure}{0.45\textwidth}
        \includegraphics[width=\textwidth]{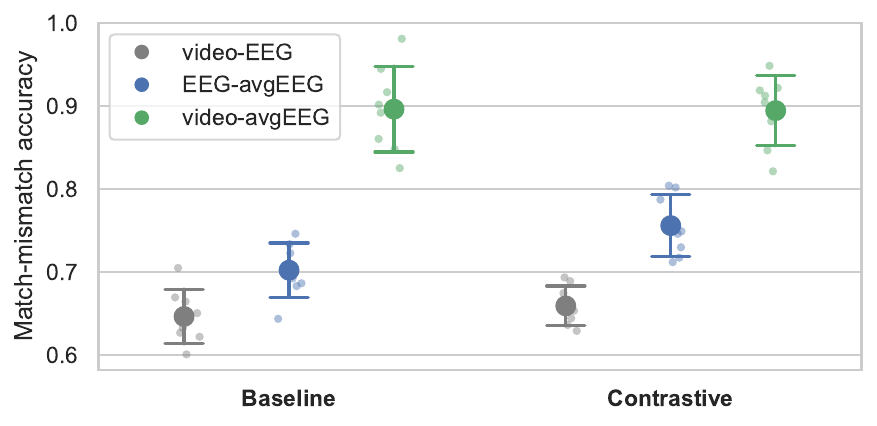}
        \caption{Under LVSO}
        \label{fig:refEEG_LVSO}
    \end{subfigure}
    \hfill
    \begin{subfigure}{0.45\textwidth}
        \includegraphics[width=\textwidth]{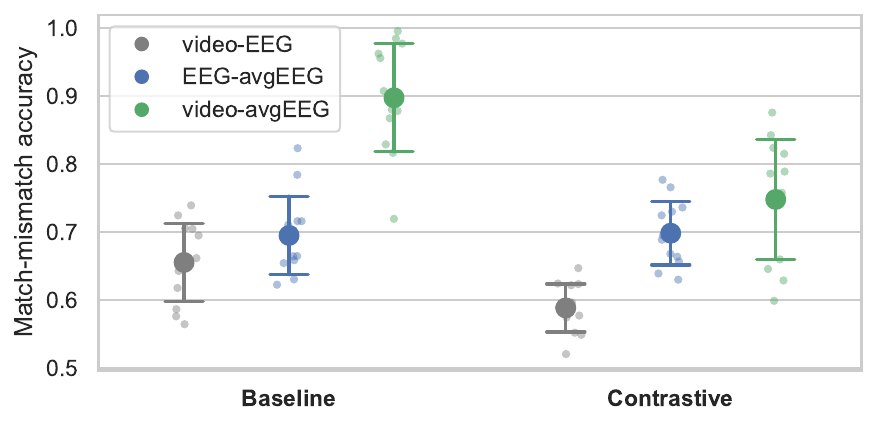}
        \caption{Under LOVO}
        \label{fig:refEEG_LOVO}
    \end{subfigure}
    \caption{MM accuracy of three tasks, video-EEG (grey), EEG-avgEEG (blue), and video-avgEEG (green), for the CCA baseline and the contrastive model, under (a) LVSO and (b) LOVO. The large dot is the mean and the error bars show $\pm1$ SD; each faint dot is one video-segment fold in (a) ($n=8$) and one video in (b) ($n=13$).}
    \label{fig:refEEG_results}
\end{figure}

 The video-avgEEG accuracy is much higher than the standard (single-subject) video-EEG accuracy in all cases (Figure~\ref{fig:refEEG_results}). Even under LOVO, where the model underperforms CCA on individual EEG, video-avgEEG reaches a mean accuracy around $75\%$ on unseen videos. Genuine, generalizable EEG-video coupling that transfers to unseen videos is therefore learned even in the regime where the model appears to fail; it is simply masked by individual noise in the EEG embedding. Moreover, video-avgEEG on held-out videos or video segments is higher than EEG-avgEEG: the video embedding matches the group EEG better than an individual subject's EEG matches it. Both comparisons point to the same conclusion, that individual noise, rather than an absence of learned alignment, is what limits the contrastive model's decodability under LVSO and LOVO.

\par This does not conflict with the strong LOSO performance of Section~\ref{sec:CL_wins_LOSO}. Individual noise limits every regime, LOSO included. What differs across regimes is not the noise, which is identical since the same subjects and EEG are used throughout, but the strength of the signal relative to that noise. For seen videos (LOSO, and partly LVSO), the model can additionally exploit stimulus-specific correspondence, whether genuine but confined to the seen content or synthesized (Section~\ref{sec:sham_experiment}). For unseen videos (LOVO), the model can only leverage generalizable video features (omitting video-specific features), which is weaker in single-subject EEG. LOSO therefore sits highest not because it escapes individual noise, but because it brings more signal to bear against the same noise floor.

\subsection{Group-Level Alignment Is Easy; Single-Subject Decoding Is Limited by Signal-to-Noise Ratio}
\label{sec:single_subject_bottleneck}

\par Prior efforts to strengthen EEG-video alignment have concentrated on finding better video representations, from hand-crafted features \cite{dmochowski2018extracting, yao2024identifying} to the pretrained VideoMAE features used here, in the hope of explaining more of the EEG variance. The group-averaged reference allows us to examine this direction separately for the shared and the individual part of the EEG. For the shared part, the video-avgEEG accuracy already reaches around $90\%$ in both LVSO and LOVO with the simple hand-crafted optical-flow feature of the linear CCA baseline, even higher than the contrastive model's video-avgEEG (around $75\%$ under LOVO; Section~\ref{sec:bottleneck}). On 30-second windows, the MM task against the group-averaged reference thus already approaches its ceiling, leaving little room to observe further gains from better video representations on this task.

\par The coupling with the individual EEG, which is the quantity that matters at deployment, is considerably weaker. The gap between the video-EEG and video-avgEEG tasks reflects a low single-subject signal-to-noise ratio (SNR) (Section~\ref{sec:bottleneck}): the stimulus-locked signal is weak relative to individual noise in a single subject. This bottleneck can be addressed from two sides. On the signal side, a more expressive video representation could capture a stronger stimulus-locked response and thereby raise the single-subject SNR. On the noise side, methods that suppress individual noise in the EEG could achieve the same, a direction that has so far received far less attention than feature design.

\subsection{Implications for Evaluation, Modeling, and Data}
\label{sec:implications}

\par The contrastive model is positioned differently against the CCA baseline under the three regimes (Sections~\ref{sec:CL_wins_LOSO}--\ref{sec:CL_loses_LOVO}), and the sham experiment (Section~\ref{sec:sham_experiment}) shows that a high LOSO accuracy can be deceptive, reflecting a synthesized correspondence rather than physically meaningful coupling. Any single regime therefore gives an incomplete, and potentially misleading, picture. This leads to what we regard as the most impactful implication of this work: models for neural tracking of natural stimuli should be evaluated under the strictest regime, holding out subjects and stimuli jointly (leave-subject-and-video-out), whenever the dataset permits it. 

\par This recommendation is not specific to video, as studied here, but applies equally to neural tracking of speech. In that domain, increasingly complex nonlinear models share the same capacity to synthesize a correspondence instead of capturing genuine neural coupling. Such models are commonly evaluated with only one axis held out, subjects or stimuli, but rarely with both held out simultaneously \cite{ciccarelli2019comparison, monesi2020lstm, accou2021modeling, puffay2023relating, thornton2022robust}, which limits the confidence that can be placed in their reported gains. Our findings clearly demonstrate the risks of using a regime that holds out only one axis and leaves the other exploitable, despite the prevailing assumption that these are rigorous evaluation schemes that are not prone to shortcuts. When a dataset does not permit joint subject-and-stimulus holdout (e.g., due to small scale, as in our case), LOSO and LOVO should at least be reported together, since the gap between them exposes how much of the apparent performance depends on having seen the stimulus or the subjects during training.

\par To improve generalization under LVSO and LOVO, we evaluated several interventions: constraining output dimensionality via low-rank temporal bases, implementing a multi-component structure, applying a multi-scale correlation loss, augmenting EEG channels, and introducing an auxiliary EEG-denoising loss. None yielded meaningful improvements over the base model. This outcome suggests two non-mutually exclusive interpretations: either the architecture requires a fundamental redesign, or the dataset inherently limits the current model's capabilities in size and/or diversity. Both scenarios point to the same underlying requirement. If the dataset is the bottleneck, scaling it is directly necessary; conversely, if a more expressive architecture is required, it will demand more data to prevent overfitting. Ultimately, future progress is gated by the size and diversity of the available data.

\par This turns attention to how such data should be collected. The fact that the model ties CCA under LVSO but underperforms under LOVO is unsurprising: the latter demands extrapolation to unseen videos, whose distribution differs from the training stimuli, and datasets in this area reuse a small set of videos, leaving a deep model little basis to generalize to novel stimuli. Broadening the stimulus set is thus likely more valuable than recording more subjects on the same videos.

\section{Conclusion}
\label{sec:conclusion}

\par We present the first study of temporal contrastive alignment between EEG and naturalistic video. Our approach aligns EEG signals with pretrained VideoMAE features using a lightweight two-branch encoder trained via a correlation-based InfoNCE objective. We benchmarked this model against a CCA baseline across three generalization regimes and a downstream selective visual attention task. Results show that the contrastive model outperforms CCA on seen stimuli, both in cross-subject (LOSO) settings and downstream attention decoding. However, it performs only on par with CCA within the stimulus manifold (LVSO) and falls behind when forced to extrapolate to entirely novel videos (LOVO).

\par Two diagnostic analyses explain this pattern. A sham experiment shows that high accuracy on seen stimuli does not certify physically meaningful EEG-video alignment: a synthesized, physically arbitrary correspondence, fit over the training stimuli, is enough to reproduce it. Yet genuine, generalizable coupling is also learned. It reaches good accuracy against a group-averaged EEG reference even on unseen videos, and is simply masked in single-subject EEG by individual noise, i.e., measurement noise and individual differences in the response to the video. Building the coupling with this clean reference is already achieved by the linear CCA baseline with a simple motion feature, which points to the SNR of single-subject EEG as the primary bottleneck.

\par These findings carry two practical messages. For applications with a fixed, known stimulus set, genuine cross-stimulus coupling is not required, and template matching with stimulus-specific features over seen stimuli already suffices. The demand for genuine, generalizable alignment becomes binding only for open-set applications that must handle novel stimuli. For the field more broadly: neural tracking of natural stimuli, whether video or speech, should be evaluated with subjects and stimuli held out simultaneously, since single-axis regimes such as LOSO or LOVO, despite their rigorous reputation, can substantially overstate genuine alignment. Further progress is gated by the size and diversity of the data. Ideally, datasets should be broadened to include more stimuli, and methods should raise the single-trial SNR, both by capturing more stimulus-locked signal and by suppressing individual noise, the latter having received far less attention so far.

\bibliographystyle{IEEEtran}
\bibliography{bstctl,strings,refs}

\end{document}